\documentclass[aps,prl,reprint,floatfix,amsmath,amssymb,twocolumn]{revtex4-2}
 \usepackage{CJK}
\usepackage{epsfig}
\usepackage{graphicx}
\usepackage{dcolumn}
\usepackage{bm}
\usepackage[colorlinks=true,citecolor=blue, dvipdfm]{hyperref}
\usepackage{amssymb}
\usepackage{enumerate}

\begin{document}
\title{Unique temporal scaling dimension for quantum criticality in open systems weakly coupled to environment}

\author{Fan Zhong}
\email{stszf@mail.sysu.edu.cn}
\affiliation{School of Physics and State Key Laboratory of Optoelectronic Materials and
Technologies, Sun Yat-sen University, Guangzhou 510275, People's Republic of China}

\date{\today}

\begin{abstract}
Probing, understanding, predicting, and controlling the real-time dynamics of quantum phase transitions in open systems are of pivotal importance to modern condensed matter physics, statistical physics, and quantum computing, among others. Here it is argued that a distinct temporal renormalization-group eigenvalue is needed for quantum criticality in open systems weakly coupled to their finite-temperature environment. This new physics enables the formulation of a general scaling theory that can accurately account for the critical properties including the specific Kibble-Zurek scaling in such open quantum systems. Remarkably, the critical exponents of time-related quantities are altered nonperturbatively regardless of how weak the coupling is, except for an Ohmic bath. Perspectives for future study are also discussed.
\end{abstract}


\maketitle

Quantum phase transitions (QPTs) describe fundamental changes in the ground state of a many-body system driven by quantum fluctuations at absolute zero~\cite{sachdev}. Unlike classical transitions, static and dynamic behaviors are inherently intertwined in QPTs. Real-time dynamics of QPTs are therefore central to modern condensed matter physics, statistical physics, and quantum computing, among others. While manipulating isolated real-time dynamics of QPTs is technically possible in some deliberately designed setups~\cite{Greiner,Kinoshita,Hofferberth}, real experiments generally inevitably involve coupling to an environment. Moreover, the absolute zero limit is thermodynamically inaccessible, making the role of finite temperature unavoidable. Therefore, probing, understanding, predicting, and controlling QPTs in open systems is of pivotal importance.

Continuous QPTs constitute one prominent class of QPTs. The associated quantum criticality shares scaling and universality as hallmarks with its classical counterpart~\cite{Mask,Cardyb,Justin,Amit}. An effective method to study such collective behavior is to drive a system through its critical point by varying the temperature or a parameter in the system Hamiltonian in time. The renormalization-group (RG) theory for linear driving was first developed for first-order phase transitions~\cite{zhongl05}, following the tradition of applying the method to study dynamics of dislocations and first-order phase transitions (see, e.g., Refs.~\cite{Ke,Zhang,Zhang1,Zhang3,zhongprl}). The theory was  later applied to classical critical phenomena~\cite{zhongxu,zhong06,Gong}. This results in a theory of finite-time scaling (FTS)~\cite{Gong,Gong1,Huang} since the rate of the driving imposes a controllable finite time scale on the evolution of the system, in close analogy to the role that the finite system size plays in the well-known finite-size scaling~\cite{fss,Brezin,Brezin1,Amit,Cardyb,Justin}. FTS enables one to describe the scaling of the entire driving process around the critical point and to detect both equilibrium and nonequilibrium critical properties. An RG theory for a nonlinear driving with multiple parameters of arbitrary form has also been developed, leading to a series of driven nonequilibrium critical phenomena such as hysteresis, negative susceptibility, and crossover between distinct regimes dominated by the different parameters in the driving~\cite{Feng}.

A driven nonequilibrium critical phenomenon is the generation of topological defects via the famous Kibble-Zurek mechanism. Originally proposed in cosmology~\cite{Kibble1,Kibble2} and later applied to condensed-matter physics~\cite{Zurek1,Zurek2}, the mechanism assumes a frozen-in diabatic regime sandwiched between two adiabatic regimes during cooling through a critical point. The resulting topological defect density is estimated from the boundary between the regimes to obey a Kibble-Zurek scaling, a power law of the cooling rate with an exponent related to both equilibrium and dynamic critical exponents. Despite extensive study~\cite{revqkz1,revqkz2,Zurek4}, several important questions remain unanswered: what the accuracy of the scaling is, how other parameters (e.g., an external field) influence it, and even whether the specific temperatures at which the defect density is estimated affect it. For instance, although the scaling is  exactly derived at the boundary, which varies with the cooling rate, the defects are sometimes directly estimated at a fixed temperature regardless of the cooling rate~\cite{Laguna1,Braun,Suzuki}. Moreover, although the topological defects themselves may be important, they are not good observables, at least in some cases. An example is the ferromagnetic Ising model where the interface between up and down spins is a topological defect. At finite temperature, the vigorous fluctuation of spins renders any meaningful estimation of topological defects unreliable. However, all these questions and others, such as how observables other than the defect density  (e.g., an order parameter) scale, are transparent in FTS, as will be seen below, together with a case in which the defect density may be an inappropriate starting point in perturbation expansions.

Returning to quantum criticality in open quantum systems, one usually considers weak coupling of the system to its environment to avoid the latter's influence on the former's properties to be studied (though, as will be demonstrated, this is fundamentally unavoidable no matter how weak the coupling is); otherwise, new phase transitions of the composite system may occur~\cite{Sudip,Weiss,Werner,Wang,Oshiyama}. The first investigation of a system driven through its own critical point at a fixed finite bath temperature yielded, via heuristic and scaling arguments, different expressions for the excitation density of coherent and incoherent contributions and a crossover between them~\cite{patane,patane1}. A systematic theoretical framework, based on the Lindblad equation~\cite{Lindblad,attal,Breuer}, was later proposed and tested using FTS. The key feature of the theory is the introduction of the coupling as a new indispensable variable, in addition to the temperature and the original scaling variables of the closed system, aiming to probe quantum criticality at finite temperature~\cite{Yin,Yin3}. The framwwork was then further validated using finite-size scaling~\cite{Yinchen,Rossini}. Recently, a specific Kibble-Zurek scaling of the excitation density~\cite{Nalbach,Kuo,Bacsi,King,Ding,Kriel} that differs from the standard one~\cite{Zurek,Polk} has been obtained at varying temperatures, across various models, driving protocols, and bath spectral densities, without resorting to finite-size scaling, partly due to technical advances in solving master equations~\cite{Prosen}. Here, I propose a general scaling theory that can systematically and accurately derive the critical properties of QPTs of systems weakly coupled to their finite-temperature environments. The central idea is that the coupling changes the time scale and its RG eigenvalue. This novel mechanism is shown to modify, in a non-perturbative but controllable manner, the critical exponents of time-related quantities and hence the universality class of the original quantum criticality, no matter how weak the coupling is, except for an Ohmic bath.

Consider a system with Hamiltonian ${\mathcal H}$ coupled to a bath, described by a Lindblad equation:~\cite{Lindblad,attal,Breuer},
\begin{equation}
	\frac{\partial \rho}{\partial t}=-i[{\cal H},\rho]+\lambda\left( [{\cal L}\rho, {\cal L}^{\dagger}]+[{\cal L}, \rho{\cal L}^{\dagger}]\right), \label{drho}
\end{equation}
where $\rho$, $t$, and $\lambda$ are the density matrix of the system, time, and coupling constant, respectively, ${\cal L}$ denotes a set of jump operators, the superscript ${\dagger}$ is the Hermitian conjugate, and square brackets represent commutators. Planck constant $\hbar$ and Boltzmann constant $k_{\rm B}$ are set to unity. The bath is initially thermalized at temperature $T$ and is characterized by a spectral density $J(\omega)=a\omega^s$ with a cutoff frequency well above the mode frequencies $\omega$, where $a$ is a dimensionless constant. The exponent $s$ specifies the bath type: $s=1$ represents an Ohmic bath, $0<s<1$ indicates a sub-Ohmic bath, while $s>1$ corresponds to a super-Ohmic bath.  

We first recapitulate the essence of the scaling hypothesis to set the stage. For a single critical point at $g=0$, the scaling hypothesis for the excitation density ${\mathfrak D}$ in a $d$-dimensional space is~\cite{Yin,Yin3}
\begin{equation}
	{\mathfrak D}(t,g,T,\lambda,L)= b^{-d}{\mathfrak D}(tb^{-z}, gb^{1/\nu}, Tb^{z},\lambda b^{(1-s)z}, L^{-1}b ),\label{sgtb}
\end{equation}
where $b$ is a length rescaling factor, $L$ the system size, and $\nu$ and $z$ are the critical exponents for the correlation length and the dynamics, respectively~\cite{Mask,Cardyb,Justin,Amit}. We have included the size of the system $L$ to incorporate finite-size effects and have neglected dimensional factors for simplicity. Equation~(\ref{sgtb}) can be derived from the RG theory in classical critical phenomena when the scaling form includes the variables $t$, $g$, $L$, and others quantities such as an order field. In the quantum context, the same equation can also be derived in the presence of $g$ and $T$, because $1/T$ plays the role of imaginary time~\cite{sachdev}. With $\hbar=k_{\rm B}=1$, frequency and energy share the same RG exponent as $T$. This determines the scaling dimension or RG eigenvalue for $\lambda$ in Eq.~(\ref{sgtb}), since $\lambda$ itself is the inverse time. The case $s=0$ has been studied previously~\cite{Yin,Yin3,Yinchen,Rossini,Bacsi}.  

The scaling hypothesis Eq.~(\ref{sgtb}) can be justified by appealing to physics. Setting $b=|g|^{-\nu}$, one finds
\begin{equation}
    {\mathfrak D}= |g|^{d\nu}f(t|g|^{\nu z}, T|g|^{-\nu z},\lambda |g|^{-(1-s)\nu z}, L^{-1}|g|^{-\nu} ),\label{sgtxi}
\end{equation}
which is a quasi-equilibrium scaling form with time and size corrections and so on, where $f(x_1,x_2,x_3,x_4)={\mathfrak D}(x_1,1,x_2,x_3,x_4)$ is a universal scaling function. The last argument of $f$ indicates that $|g|^{-\nu}$ is a length scale. Indeed, it is simply the correlation length $\xi\sim|g|^{-\nu}$. Accordingly, for a given $g$ and a certain value of ${\mathfrak D}$, the first argument of $f$ correctly produces the corresponding correlation time $\zeta_{\rm corr}\sim|g|^{-\nu z}\sim\xi^z$ and its defining dynamic critical exponent. Moreover, the leading behavior of Eq.~(\ref{sgtxi}) becomes asymptotically ${\mathfrak D}\sim\xi^{-d}$, which is the conventional definition of the topological defect density. This asymptotic behavior is reached when all scaled variables are assumed to be vanishingly small. For example, $L^{-1}|g|^{-\nu}\ll1$, i.e., $\xi\ll L$, correctly quantifies the thermodynamic limit. In addition, $\lambda|g|^{-(1-s)\nu z}\ll1$ and $T|g|^{-\nu z}\ll1$ are equivalent to $\lambda\ll\xi^{-z}\sim\zeta_{\rm corr}^{-1}$ (for $s=0$) and $T\ll\Delta$, respectively, where $\Delta\sim|g|^{\nu z}\sim\xi^{-z}$ is the energy gap above the ground state. These two inequalities indicate that $\lambda$ and $T$, which are associated with dephasing and thermal excitation, become negligible when sufficiently close to the quantum critical point.

Now we consider real-time driving across the quantum critical point by nonlinearly ramping a control parameter, for example,
\begin{equation}
g=Rt^n,\label{grt}
\end{equation}
where $n$ is a constant exponent and $R>0$ is the rate of the ramp. Let $r$ be the RG eigenvalue for $R$, implying that at scale $b$, $R$ becomes $Rb^{r}$. The RG theory predicts that~\cite{Feng}
\begin{equation}
	r=nz+1/\nu.\label{rznu}
\end{equation}  
Since only two out of the trio $g$, $R$, and $t$ are independent, we replace $t$ by $R$ in Eq.~(\ref{sgtb}) and set $b=R^{-1/r}$, yielding an FTS scaling form~\cite{Gong,Gong1,Huang},
\begin{equation}
	{\mathfrak D}= R^{d/r}f_R(gR^{-1/r\nu}, TR^{-z/r},\lambda R^{-(1-s)z/r}, L^{-1}R^{-1/r} ),\label{sgtr}
\end{equation} 
with finite-size corrections, where $f_R$ is another scaling function. Throughout, exponents such as $1/r\nu$ are understood as $1/(r\nu)$ to reduce parentheses. The leading term $R^{d/r}$ in Eq.~(\ref{sgtr}) is simply the standard Kibble-Zurek scaling, which can only be accurately achieved when all the scaled variables in $f_R$ are either constant or negligible. In particular, the condition $\hat{g}R^{-1/r}=1$ precisely defines the frozen value $\hat{g}$~\cite{Huang}. So, if $g$ is constant (as opposed to $\hat{g}$, which varies with $R$), then $f_R$ still depends on $R$. As a result, ${\mathfrak D}$ is not simply proportional to $R^{d/r}$, even if all remaining variables can be neglected. In addition, it is clear that $R^{-z/r}$ is a controllable driving time scale, which is why Eq.~(\ref{sgtr}) is called the FTS form.

In the standard Kibble-Zurek scaling derived from the leading behavior in Eq.~(\ref{sgtr}), the dissipation term involving $\lambda$ acts only as a perturbation, i.e., $\lambda R^{-(1-s)z/r}\ll1$, or equivalently, $\lambda^{-1}\gg R^{-(1-s)z/r}$. This implies that the dissipation time is much longer than the driving time and can therefore be neglected. To find a Kibble-Zurek scaling that is proportional to $\lambda$, we can expand the scaling function $f_R$ with respect to this variable, resulting in
\begin{equation}
	{\mathfrak D}(R,g,T,\lambda,L)={\mathfrak D}_{\rm coh}+{\mathfrak D}_{\rm inc},\label{sgtkz}
\end{equation}
where the coherent contribution to the excitation density is
\begin{eqnarray}
   {\mathfrak D}_{\rm coh}\sim R^{\sigma_{\rm coh}},\qquad\quad\label{ncoh}\\
   \sigma_{\rm coh}=d/r=d\nu/(1+n\nu z),\label{scoh} 
\end{eqnarray}
and the incoherent one is ${\mathfrak D}_{\rm inc}\sim \lambda R^{\sigma_{\rm inc}}$ with the Kibble-Zurek scaling exponent $\sigma_{\rm inc}=[d-(1-s)z]/r=[d-(1-s)z]\nu/(1+n\nu z)$. Equation~(\ref{sgtkz}) explicitly justifies the separation of the two contributions to ${\mathfrak D}$, while Eq.~(\ref{scoh}) for $\sigma_{\rm coh}$ under linear driving ($n=1$) is the standard Kibble-Zurek scaling~\cite{Zurek,Polk} .
For the $d=1$ QPTs in the transverse Ising model and in the Kitaev model, where $\nu=z=1$, we find $\sigma_{\rm inc}=sz/(1+n)$, different from the standard coherent exponent $\sigma_{\rm coh}=1/(1+n)$. Yet, this $\sigma_{\rm inc}$ has no physical counterpart, except for $s=1$, where $\lambda$ is dimensionless (see below).

To obtain a better estimate of $\sigma_{\rm inc}$, we can expand the correlation length $\xi$ and employ ${\mathfrak D}\sim\xi^{-d}$. The FTS form for $\xi$~\cite{Zuo} is analogue to that for ${\mathfrak D}$ given in Eq.~(\ref{sgtr}); the only modification is the replacement of the leading asymptotic behavior by $R^{-1/r}$, a driving length scale. Therefore
\begin{eqnarray}
	{\mathfrak D}_{\rm inc}\sim \lambda^{-d}R^{\sigma_{\rm inc}},\qquad\qquad\quad\label{ninc}\\
	\sigma_{\rm inc}=\frac{d[1+(1-s)z]}{r}=\frac{d\nu[1+(1-s)z]}{(1+n\nu z)}.\label{sinc} 
\end{eqnarray}
$\sigma_{\rm inc}=(2-s)/(1+n)$ for the two specified models. Equation~(\ref{sinc}) agrees with numerical results for $s=0$ and $n\nu=1$~\cite{Bacsi} [cf. Eq.~(\ref{sg}) below], as well as for the exceptional case of $s=1$ mentioned above. Also consistent is the $\lambda^{-d}$ factor in Eq.~(\ref{ninc})~\cite{Bacsi}, which differs from the linear factor in the direct expansion of Eq.~(\ref{sgtr}). It is this factor that renders the direct expansion of ${\mathfrak D}$ incorrect. Note that the expressions for $\sigma_{\rm inc}$ and $\sigma_{\rm coh}$ differ, as seen from Eqs.~(\ref{scoh}) and~(\ref{sinc}). Their crossover occurs at $\lambda R^{-z/r}\approx 1$ for $s=0$, when the two time scales involved, the dephasing time $\lambda^{-1}$ and the driving time $R^{-z/r}$, become comparable. The above result illustrates that the density of the topological defects may be an unsuitable quantity to start with.

The result that the perturbation expansion produces a correct outcome only for the two special cases of $s=0$ and $s=1$ is profound. As can be seen from Eq.~(\ref{sgtb}), at $s=0$, $\lambda$ possesses the same scaling dimension as $t^{-1}$, while at $s=1$, $\lambda$ is dimensionless and serves purely as a time unit. Conversely, the fact that the standard time scale $b^{-z}$ fails for $s\neq 0, 1$ suggests that it is inapplicable. The key insight is therefore to adopt $\lambda^{-1} b^{-sz}$ as the time scale whenever it dominates over $b^{-z}$. This is similar to the case of critical phenomena with memory~\cite{Zeng,Zeng1}. When the decay exponent of the long-range temporal interaction is smaller than $1$, a value which originates from the first-order time derivative in the purely dissipative Langevin equation, the time scale is then determined by the long-range temporal interaction rather than the standard time-derivative term. Note that $\lambda$ serves only as a time unit, in analogy to the kinetic coefficient in the Langevin equation for critical dynamics~\cite{Hohenberg,Folk}. Therefore, Eq.~(\ref{sgtb}) is modified to
\begin{equation}
	{\mathfrak D}(t,g,T,\lambda)= b^{-d}{\mathfrak D}(t\lambda^{-1} b^{-sz}, gb^{1/\nu}, Tb^{z}, L^{-1}b^{-1}),\label{sgtbr}
\end{equation}
All variables other than $t$ retain their original RG eigenvalues. This seems to be in line with the expectation that the bath is only a weak perturbation which causes minimal alteration of the original quantum criticality~\cite{Yin3}. However, we will shortly discover surprises and will return to this later in the discussion. Note that Eq.~(\ref{sgtbr}) is essentially Eq.~(\ref{sgtb}) in the special Ohmic case $s=1$. However, we will see that even in this case, non-perturbative effects emerge.

Next, we extract predictions from our central scaling hypothesis, Eq.~(\ref{sgtbr}). A first direct result can be readily reached by setting $b=(t/\lambda)^{1/sz}$, leading to
\begin{equation}
	{\mathfrak D}(t,g,T,\lambda)= (t/\lambda)^{-d/sz}f_t(g(t/\lambda)^{1/s\nu z}, T(t/\lambda)^{1/s}),\label{sgt}
\end{equation}
where $f_t$ is yet another scaling function and we have dropped the system size for clarity, although it can be easily included if needed. Equation~(\ref{sgt}) indicates that ${\mathfrak D}$ decays as $(t/\lambda)^{-d/sz}$ exactly at the quantum critical point $g=T=0$, in agreement with the extant results~\cite{Ding}. Note that in the standard setup of Eq.~(\ref{sgtb}), setting $b=t^{1/z}$ results in the standard coherent evolution ${\mathfrak D}\sim t^{-d/z}$, differing by the power of bath characteristic $s$. The two different decay forms arise from distinct sources and are therefore independent. They differ only in the time span over which they persist, which depends on $s$~\cite{Ding}. Away from the critical point, the validity of Eq.~(\ref{sgt}) requires $g(t/\lambda)^{1/s\nu z}\ll1$ and $T(t/\lambda)^{1/s}\ll1$, or, $t\ll\lambda|g|^{-svz}\sim\zeta_{\rm corr}$ and $t\ll\lambda T^{-s}$. These two expressions are reasonable because they state that the time must be much smaller than the equilibrium correlation time $\zeta_{\rm corr}$ and the imaginary time $T^{-s}$; otherwise the system would have nearly equilibrated. What may appear surprising is that the definitions of both the correlation time and imaginary time now have an extra $s$ power. However, these are nothing more than natural as the time dimension itself simply carries that extra power, as is evident from Eq.~(\ref{sgtbr}). In fact, such time scales have been employed to heuristically derive the specific Kibble-Zurek scaling~\cite{Bacsi,Kuo,Ding}.

We proceed to consider driving. We can ramp either $g$ as Eq.~(\ref{grt}) or $T$ as
\begin{equation}
	T=\pm R_Tt^m\label{Trt}
\end{equation}
with constant $R_T>0$ and $m$, where $\pm$ applies to heating or cooling. Now $gb^{1/\nu}=Rb^{r_g} t^n\lambda^{-n}b^{-nsz}$ from Eqs.~(\ref{grt}) and~(\ref{sgtbr}) and a similar equation for $T$, giving rise to
\begin{equation}
	r_g=nsz+1/\nu,\qquad r_T=(1+ms)z.\label{rgnsz}
\end{equation}
Again replacing $t$ with $R$ or $R_T$ transforms Eq.~(\ref{sgtbr}) into
\begin{eqnarray}
	{\mathfrak D}(R,g,T,\lambda)= b^{-d}{\mathfrak D}(\lambda^{-n}Rb^{r_g}, gb^{1/\nu}, Tb^{z}),~\label{sgtbrr}\\
	{\mathfrak D}(R,g,T,\lambda)= b^{-d}{\mathfrak D}(\lambda^{-m}R_Tb^{r_T}, gb^{1/\nu}, Tb^{z}).\label{sgtbrT}
\end{eqnarray} 
Equation~(\ref{sgtbrT}) is essentially the homogeneous function found in Ref.~\cite{King} for a linear driving $m=1$ through a specific model and the resulting exact rate equation. Here we arrive at it generally without such restrictions. Setting $b=(\lambda^{-n}R)^{-1/r_g}$ or $b=(\lambda^{-m}R_T)^{-1/r_T}$ then yields the FTS forms for the two driving protocols as
\begin{eqnarray}
	{\mathfrak D} =(\lambda^{-n}R)^{d/r_g}f_g(g\lambda^{-n}R^{-1/r_g\nu}, T\lambda^{-n}R^{-z/r_g}),\qquad~\label{sgrr}\\
	{\mathfrak D} =(\lambda^{-m}R_T)^{d/r_T}f_T(g\lambda^{-m}R^{-1/r_T\nu}, T\lambda^{-m}R^{-z/r_T}),\quad~~\label{sgTr}	
\end{eqnarray} 
where $f_g$ and $f_T$ are universal scaling functions. Consequently, applying Eq.~(\ref{rgnsz}), we arrive at the Kibble-Zurek scaling exponents as
\begin{equation}
	\sigma_g=d\nu/(1+nsz\nu),\quad \sigma_T=d/(1+ms)z,\label{sg}
\end{equation}
both of which exactly match the results in Ref.~\cite{Bacsi,Kriel}. Therefore, the FTS theory developed here explains the specific Kibble-Zurek scaling accurately, in the sense that not only the scaling exponent itself, but also other contributions, whether relevant or irrelevant in the RG sense, are all adequately treated on an equal footing. 

We note that the functional form of $\sigma_g$ differs from that of $\sigma_{\rm inc}$, Eq.~(\ref{sinc}), except for the Ohmic case $s=1$. Their values happen to coincide when $s=0$ and $n\nu=1$, which is satisfied, for instance, for linear driving through a quantum critical point with $\nu=1$. However, even for the Ohmic case, the dependence on $\lambda$ is different for nonlinear driving, see Eqs.~(\ref{ninc}) and~(\ref{sgrr}). These observations imply that the coupling cannot, in general, be regarded as a perturbation, however weak it is, even though it is initially designed to be sufficiently weak so that the critical properties of the original QPT would not be influenced. This influence is corroborated by the distinct exponents of time-related quantities such as the correlation time and $T$, as exemplified by the consequences derived from Eq.~(\ref{sgt}). From the perspective of universality class, this change in exponents signifies a change of universality class with varying $s$, except for $s=1$. Nevertheless, from the standpoint of determining the universality class of the original QPT, it can be straightforwardly recovered from the relations among the exponents once $s$ is known.

In summary, we have revealed a novel mechanism underlying the quantum criticality of open systems weakly coupled to their finite-temperature environment: the time scale and its RG eigenvalue are modified by the coupling. Consequently, a systematic scaling theory has been developed, which accurately explains the observed special Kibble-Zurek scaling, capturing not only the scaling exponent itself, but also other contributions that are either relevant or irrelevant in the RG sense. The theory shows that, for a non-Ohmic bath, the critical exponents of the time-related quantities such as the correlation time and the temperature are also modified. As a result, the universality class is also changed regardless of how weak the coupling is, in contrast to initial expectations. Nonetheless, the critical exponents of the original QPT can still be derived from the bath characteristic constant $s$ on which they depend. In addition, we have demonstrated that, in the Ohmic case, the Kibble-Zurek scaling exponent is correctly captured by a perturbative expansion of the correlation length, but not by that of the excitation density itself. Nevertheless, under nonlinear driving, the dependence of the excitation density on the coupling deviates from the non-perturbative prediction. This reveals a deficiency in using the defect density as the starting point of perturbative expansions, thereby highlighting the necessity of the non-perturbative theory.

Several intriguing questions arise for future study. We have seen that the new time scale modifies the critical exponents of some quantities. However, in the critical phenomena with memory~\cite{Zeng,Zeng1}, the long-range temporal interaction of the system radically alters its critical behavior, even at the mean-field level, by introducing a series of brand new phenomena such as violation of scaling laws, spatiotemporal dimension shifts, discontinuous crossover from short-range to long-range behavior, anomalous system-size dimensions, and peculiar behaviors above the upper critical dimension, in sharp contrast to the case of long-range spatial interaction~\cite{Fisher,Sak}. It is then desirable to know whether such behaviors can emerge here in the open quantum systems. Another question concerns the universality class. It has been shown that the coupling can lead to a new dissipative phase transition~\cite{Sudip,Werner,Oshiyama}. We have shown above that the weak coupling results in new universality classes in which some quantities exhibit distinct critical exponents. A natural question to pose is then whether the two transitions are related. Yet another question is related to new exponents in the driven weakly-coupled open quantum systems we consider. Although we have new critical exponents from the coupling, they are all determined by the known equilibrium and dynamic critical exponents, as well as $s$. The question here is whether fundamentally new exponents different from the foregoing ones can emerge in such driven open quantum systems. Some new exponents have been found by breaking self-similarity symmetry determined by $L^{-1}R^{-1/r}$ as appears in Eq.~(\ref{sgtr})~\cite{Yuan,Yuan1,Yuan2}: If $L^{-1}R^{-1/r}$ is fixed such that systems of different sizes contain the same number of driving lengths $R^{-1/r}$, i.e., self-similarity symmetry holds, then no new exponents are needed; whereas if it is broken, new exponents must be invoked. Note that these new exponents stem from critical phenomena rather than phase ordering~\cite{Biroli,Jeong}.

\begin{acknowledgments}
This work was supported by National Natural Science Foundation of China (Grant No. 12175316).
\end{acknowledgments}

\end{document}